\documentclass[12pt]{article}
\usepackage{epsfig,amsfonts,amssymb,amsbsy,amsbsy,array}
\renewcommand{\theequation}{\thesection.\arabic{equation}}
\def\cvp{\raise 2pt\hbox{,}}

\def\tr{\mathop{\rm tr}\nolimits}
\def\im{\mathop{\rm Im}\nolimits}
\def\re{\mathop{\rm Re}\nolimits}

\def\d{{\rm d}}
\def\suN{{\rm SU}(N)}
\def\uN{{\rm U}(N)}

\def\wl{W_{\rm low}}\def\wt{W_{\rm tree}}\def\weff{W_{\rm eff}}
\def\ww{{\cal W}}
\def\La{\Lambda}\def\Lu{\Lambda_{\rm u}}
\def\la{\lambda}\def\lc{\lambda_{\rm c}}\def\lck{\lambda_{{\rm c},k}}
\def\p{|p\rangle}\def\q{|q\rangle}

\def\plb#1#2#3{{\it Phys.\ Lett.\ }{\bf B #1} (#2) #3}
\def\npb#1#2#3{{\it Nucl.\ Phys.\ }{\bf B #1} (#2) #3}

\def\prl#1#2#3{{\it Phys.\ Rev.\ Lett.\ }{\bf #1} (#2) #3}
\def\jhep#1#2#3{{\it J. High Energy Phys.\ }{\bf #1} (#2) #3}

\def\pr#1#2#3{{\it Phys.\ Rep.\ }{\bf #1} (#2) #3}

\begin{document}
%
%
\pagestyle{empty}
{\parskip 0in

\hfill NEIP-02-008

\hfill LPTENS-02/49

\hfill hep-th/0211069}

\vfill
\begin{center}
{\LARGE Quantum parameter space and double scaling limits}

\medskip

{\LARGE in ${\cal N}=1$ super Yang-Mills theory}

\vspace{0.4in}

Frank F{\scshape errari}{\renewcommand{\thefootnote}{$\!\!\dagger$}
\footnote{On leave of absence from Centre 
National de la Recherche Scientifique, Laboratoire de Physique 
Th\'eorique de l'\'Ecole Normale Sup\'erieure, Paris, France.}}
\\
\medskip
{\it Institut de Physique, Universit\'e de Neuch\^atel\\
rue A.-L.~Br\'eguet 1, CH-2000 Neuch\^atel, Switzerland}\\
\smallskip
{\tt frank.ferrari@unine.ch}
\end{center}
\vfill\noindent
We study the physics of ${\cal N}=1$ super Yang-Mills theory with
gauge group $\uN$ and one adjoint Higgs field, by using the recently
derived exact effective superpotentials. Interesting phenomena occur
for some special values of the Higgs potential couplings. We find
critical points with massless glueballs and/or massless monopoles,
confinement without a mass gap, and tensionless domain walls. We
describe the transitions between regimes with different patterns of
gauge symmetry breaking, or, in the matrix model language, between
solutions with a different number of cuts. The standard large $N$
expansion is singular near the critical points, with domain walls
tensions scaling as a fractional power of $N$. We argue that the
critical points are four dimensional analogues of the Kazakov critical
points that are commonly found in low dimensional matrix integrals. We
define a double scaling limit that yields the exact tension of BPS
two-branes in the resulting ${\cal N}=1$, four dimensional
non-critical string theory. D-brane states can be deformed
continuously into closed string solitonic states and vice-versa along
paths that go over regions where the string coupling is strong.

\vfill

\medskip
%
\begin{flushleft}
October 2002
\end{flushleft}
\newpage\pagestyle{plain}
\baselineskip 16pt
\setcounter{footnote}{0}

%
\section{The big picture}

In a series of papers \cite{F1,F2,F3,F2D,Fnp}, that are reviewed in
\cite{Frevue}, an attempt to generalize the matrix model approach to
non-critical strings \cite{MM} to the case of four dimensional
theories has been made. The basic idea is to replace the simple low 
dimensional matrix integrals
\begin{equation}
\label{MMint}
\int\!\d^{N^{2}}\! M\, e^{-N\tr V(M;g_{j})}
\end{equation}
used in \cite{MM} by four dimensional gauge theory path integrals 
with adjoint Higgs fields,
\begin{equation}
\label{MM4D}
\int\! \left[ \d M(x)\strut\right]\,
\exp\left[ -N\int\!\d^{4}x\, {\cal L}_{\rm YM}(M,\partial M)\right] ,
\end{equation}
where $M$ represents a collection of $N\times N$ 
hermitian matrices including the 
four components of the vector potential as well as the Higgs fields. 
The parameters in the potential $V(M;g_{j})$ appearing in (\ref{MMint}),
which are 
adjusted in the classic approach to critical values for which the 
large $N$ expansion of the matrix integral (\ref{MMint}) breaks down, 
are replaced in the four dimensional case by Higgs vacuum expectation 
values (which can be moduli), or more generally by the Higgs couplings 
appearing in the Higgs potential. It was argued in 
\cite{F1,F2,F3,F2D,Fnp} that a non-trivial low energy physics 
generically develops for some special values of these couplings. The 
large $N$ expansion then suffers from IR divergences. Those IR 
divergences are very specific, and can be compensated for by taking 
$N\rightarrow\infty$ and approaching the critical points in a 
correlated way. The resulting double scaled theories are conjectured 
to be string theories, whose continuous world-sheets are constructed 
from the large 't Hooft diagrams of the parent gauge theory.
One of the highlight of this approach is that
the double scaling limits provide full non-perturbative definitions 
of the corresponding string theories. This stems from the fact that 
the supersymmetric gauge theory path integrals are non-perturbatively 
defined for all values of the parameters, unlike the simple 
integrals (\ref{MMint}) which suffer from instabilities when the 
potential $V$ is unbounded from below.

The above picture has been tested quantitatively on the moduli space of
${\cal N}=2$ supersymmetric gauge theories \cite{F2,F3}. Moreover, it was
argued in \cite{F1} by studying two-dimensional toy models that the main
features do not depend on supersymmetry, as long as one replaces the moduli
space by the parameter space of Higgs couplings. In the present paper, we
propose to test the validity of those general ideas in the context of
${\cal N}=1$ supersymmetric gauge theories. This was motivated by the
recent progresses made in the calculation of the effective superpotentials
\cite{CIV,DV,fer}. We are going to compute the quantum corrections to the
classical space of parameters, and we indeed find qualitative similarities
with the quantum moduli space of ${\cal N}=2$ theories. This fits well with
the ideas advocated in \cite{F1}. We will also discover interesting new
aspects of ${\cal N}=1$ gauge theories.

We focus on the simplest possible examples, based on the
$\uN$ theories with a single adjoint Higgs particle. The elementary
fields of the model are the ${\cal N}=1$ vector multiplet $V$ or its
associated field strenght $W_{\alpha} = -\bar D^{2} e^{-2V}D_{\alpha}
e^{2V} /8$, that contains the gauge fields $A_{\mu}$ and the gluinos
$\psi$, and a chiral multiplet $\Phi$ in the adjoint representation
whose lowest component $\phi$ is the complex Higgs field.
The lagrangian is
\begin{equation}
\label{lag}
{\cal L} = {1\over 4\pi}\im\tr\tau_{\rm YM}\left[\int\!\d^{2}\theta\,
W^{\alpha}W_{\alpha} + 2\int\!\d^{2}\theta\d^{2}\bar\theta\, 
\Phi^{\dagger} e^{2V}\Phi\right] + 
2N\re\tr\!\int\!\d^{2}\theta\, \wt(\Phi)\, ,
\end{equation}
with a complexified bare gauge coupling constant
$\tau_{\rm YM} = \theta /(2\pi) + 4i\pi/g^{2}_{\rm YM}$ and a tree-level 
superpotential $\wt(\Phi)$. 
Quantum mechanically, the complexified gauge coupling is
replaced by a complexified mass scale $\Lu$ such that
\begin{equation}
\label{RG}
\tau_{\rm YM} = {iN\over\pi} \ln {\Lambda_{0}\over\Lu}\,\cvp
\end{equation}
where $\Lambda_{0}$ is the UV cut-off.
For the most part of the paper we consider
\begin{equation}
\label{Wtree}
\wt(\Phi) = {m\over 2}\, \Phi^{2} + {g\over 3}\, \Phi^{3}
\end{equation}
with a non-zero cubic coupling $g$. Classically, the gauge group either
is unbroken when $\langle\phi\rangle_{\rm cl} =0$ or
$\langle\phi\rangle_{\rm cl} =-m/g$, or can be broken down to ${\rm
U}(N_{1})\times {\rm U}(N-N_{1})$ when $N_{1}$ eigenvalues of $\phi$
are chosen to be $0$ and $N-N_{1}=N_{2}$ are chosen to be $-m/g$. When
$m=0$, the Higgs field becomes critical. This is a singularity on the
classical space of parameters, through which regimes with different
patterns of gauge symmetry breaking are connected. We will see that
this simple picture is modified in an interesting and subtle way in
the quantum theory.

In Section 2, we discuss the physics of the phases with unbroken 
gauge group, by using in particular the field theoretic exact 
superpotentials derived in \cite{fer}. The standard lore is that when
$|m|\gg|\Lu|$, the Higgs field can be integrated out, and the low 
energy physics is governed by pure $\uN$ super Yang-Mills.
This theory has a
running gauge coupling characterized by a low energy mass scale $\La$
which is related to $\Lu$ and $m$ by a standard matching relation,
\begin{equation}
\label{ll}
\La^{3} = m\Lu^{2} \, .
\end{equation}
Pure super Yang-Mills confines and develops a mass gap of order $|\La|$. 
The chiral symmetry ${\mathbb Z}_{2N}$ is spontaneously broken to 
${\mathbb Z}_{2}$ by the 
gluino condensate $\langle\tr\psi^{2}\rangle$ which in the $k^{\rm th}$
vacuum is proportional to $N\La^{3} e^{2i\pi k/N}$. There are
domain walls connecting the $N$ different vacua. The tensions of the 
lightest domain walls go like $N|\La^{3}|$ at large $N$, consistently 
with a D-brane interpretation \cite{Wbrane}. This simple 
picture is significantly changed by the Higgs field self-interactions.
When the dimensionless combination of couplings
\begin{equation}
\label{lamdef}
\la ={8 g^{2}\Lu^{2}\over m^{2}}= {8 g^{2}\La^{3}\over m^{3}}
\end{equation}
goes to any of the $N$ critical values
\begin{equation}
\label{critlam}
\lck = e^{-2i\pi k/N},\quad 0\leq k\leq N-1\, ,
\end{equation}
we have a phase with a massless glueball and a tensionless domain wall. By
calculating explicitly the monopole condensates, we show that we can
have confinement without a mass gap. Moreover, the critical points are
branching points on the space of parameters, modifying drastically the
classical structure. It turns out that some components of the parameter
space corresponding to the classical vacua $\langle\phi\rangle_{\rm cl} 
=0 $ and $\langle\phi\rangle_{\rm cl} =-m/g$ 
are actually glued together along branch cuts.

In Section 3, we adopt a more geometrical point of view and
provide a general discussion of the quantum 
space of parameters. By analysing the phases with a broken gauge group, 
which are described by the two-cut solution of the matrix model \cite{DV},
we show that there is an extremely rich structure, with connections 
with the unbroken phases through massless monopole points. For example, 
if $N$ is even, the singular points (\ref{critlam}) correspond to
contact points between the $\uN$ and ${\rm U}(N/2)\times {\rm U}(N/2)$ 
phases. From the geometrical point
of view, the transitions connect solutions with a different number 
of cuts. A technical analysis of the multi-cut matrix model, 
including the derivation of results used in the main text, is 
included in the Appendix. 

Section 4 is devoted to the study of the large $N$ limit. A striking 
feature is that the large $N$ expansion breaks down at the 
singularities.
This is proven explicitly for the critical points (\ref{critlam}), by
calculating for example the exact
tension of the domain walls and expanding at large $N$. We can show that in
the double scaling limit
\begin{equation}
\label{dsca}
\la\rightarrow\lc\, , \quad N\rightarrow\infty\, ,\quad N(\lc -\la) = {\rm 
constant} = 1/\kappa\, ,
\end{equation}
the renormalized (in the world-sheet sense) tensions of the domain 
walls have well-defined limits that are interpreted as giving the 
exact tensions of two-branes in the resulting non-critical string 
theory. An interesting aspect is that it is possible to deform  
a D-brane continuously into a solitonic brane and vice-versa, by going over regions 
where the string coupling is strong. A similar result holds before scaling
for the domain walls of the original gauge theory.
We will also briefly mention multicritical points.

\section{The phases with unbroken gauge group}
\setcounter{equation}{0}
\subsection{Exact superpotentials}

Our basic tool in the $\uN$ vacua is the exact quantum effective
superpotential, discussed for example in \cite{fer},
\begin{equation}
\label{wquantum}
W_{\rm q} = \sqrt{2}\sum_{m=1}^{N-1}
\tilde M_{m} M_{m} A_{D,m}(u_{p},\Lu) +  mu_{2}+gu_{3} \, ,
\end{equation}
where $u_{p} = \tr\Phi^{p}/p$, $M_{m}$ and $\tilde M_{m}$ 
are monopole fields coupling to $N-1$ magnetic ${\rm U}(1)$ gauge 
fields, and the $A_{D,m}$ are known functions of the $u_{p}$ \cite{SW}.
We will mainly use the effective superpotential $\weff$ 
for the fields $z$ and $S$ defined by
\begin{equation}
\label{deffield}
z = {u_{1}\over N} = {\tr\Phi\over N}\,\cvp\qquad
S = -{\tr W^{\alpha}W_{\alpha}\over 16N\pi^{2}}\,\cdotp
\end{equation}
The normalizations are chosen such that the fields are of order one at
large $N$. The operator $S$ is the glueball chiral superfield whose lowest
component is proportional to the gluino bilinear $\tr\psi^{2}$. It was
explained in \cite{fer} how to derive $\weff (z,S)$ from (\ref{wquantum}),
and the result is (see equation (24) of \cite{fer}, with the slightly
different convention that $\La$ in \cite{fer} is $\Lu$ presently)
\begin{equation}
\label{exW}
\weff (z,S) = {Nmz^{2}\over 2} + {Ngz^{3}\over 3} + S\ln 
\left[ {e\La^{3} \left( \strut 1+2gz/m\right)\over S}\right]^{N} .
\end{equation}
The fields $z$ and $S$ have generically masses of 
order $|\La|$. The 1PI superpotential (\ref{exW}) can be used to 
calculate the exact expectation values $\langle z\rangle$ and $\langle 
S\rangle$ upon extremization. We will show 
that for some special values of the parameters, the fields $z$ or
$S$ actually become critical. In that case, it can be useful to think of
(\ref{exW}) as a low energy superpotential. The derivatives of 
$\weff$ take very simple forms,
\begin{eqnarray}
&& \partial_{S}\weff = 
\ln\Biggl[ {\La^{3} \left(\strut 1+2gz/m\right)\over S}\Biggr]^{N}\, 
,\label{dSW}\\
&& \partial_{z}\weff = N(m z + g z^{2}) + {2 g N S\over m+2gz}\,\cdotp
\label{dzW}
\end{eqnarray}
Equation (\ref{dSW}) can be used to integrate out $S$, which yields the 
effective superpotential for $z$ in the $k^{\rm th}$ vacuum,
\begin{equation}
\label{wz}
\ww^{(k)}(z) = N\La^{3}e^{2i\pi k/N} + {2 Ng\La^{3}e^{2i\pi k/N}z\over m} + 
{Nmz^{2}\over 2} + {Ngz^{3}\over 3}\,\cvp\quad  0\leq k\leq N-1 .
\end{equation}
In a similar way, equation (\ref{dzW}) can be used to integrate out $z$, 
which yields the effective superpotential for the glueball superfield $S$,
\begin{equation}
\label{wS}
W(S) = {NmZ(S)^{2}\over 2} + {NgZ(S)^{3}\over 3} + S\ln 
\left[ {e\La^{3} \left( \strut 1+2gZ(S)/m\right)\over S}\right]^{N} .
\end{equation}
The function $Z(S)$ is determined by the equation
\begin{equation}
\label{Zdef}
2gS = -Z(m+gZ)(m+2gZ)
\end{equation}
and the classical limit $\lim_{S\rightarrow 0}Z=0$ or $\lim_{S\rightarrow
0}Z=-m/g$ depending on the vacua one considers. It
is simpler to use the derivative of $W(S)$, which is expressed in
terms of $\delta = S/\left(\strut \La^{3} (1+2gz/m)\right)$ as
\begin{equation}
\label{wp}
W'(S) = -\ln\delta^{N}\, .
\end{equation}
Using the dimensionless variable defined in (\ref{lamdef}),
$\delta$ is determined by the requirements
\begin{equation}
\label{deltaeq}
\delta^{2} ( 1 - \la\delta) = S^{2}/\La^{6}\, ,\quad \delta = \pm S/\La^{3} + 
{\cal O}(\la)\, ,
\end{equation}
the plus sign corresponding to the vacua with $\langle\phi\rangle_{\rm 
cl} = 0$ and the minus sign to $\langle\phi\rangle_{\rm cl} = -m/g$.
Interestingly, $W(S)$ given by (\ref{wS}) and (\ref{Zdef}) can be
identified with the sum of planar diagrams in the one matrix model
\cite{DV}, as was explicitly checked in \cite{fer}, and as is further 
discussed in the Appendix.

The expectation values for $S$ and $z$ in the vacuum $|k\rangle$
(associated with $\langle\phi\rangle_{\rm cl} = 0$) or the vacuum
$|k\rangle '$ (associated with $\langle\phi\rangle_{\rm cl} = -m/g$),
$0\leq k\leq N-1$, are straightforwardly deduced from the above
equations,
\begin{eqnarray}
&&\hskip -1cm \langle S\rangle_{k} = \La^{3}e^{2i\pi k/N}
\sqrt{\displaystyle 1-\la e^{2i\pi k/N}} \, ,\quad
\langle z\rangle_{k} = -{m\over 2g} \Bigl( 1-\sqrt{\displaystyle 1-\la 
e^{2i\pi k/N}}\Bigr) \, ,\nonumber\\
&&\hskip -1cm\langle S\rangle^{'}_{k} = -\La^{3}e^{2i\pi k/N}
\sqrt{\displaystyle 1-\la e^{2i\pi k/N}} \, ,\quad\!\!
\langle z\rangle^{'}_{k} = -{m\over 2g} \Bigl( 1+\sqrt{\displaystyle 1-\la 
e^{2i\pi k/N}}\Bigr) \, .\label{Szvev}
\end{eqnarray}
By replacing $z$ in (\ref{wz}) (or $S$ in (\ref{wS})) by its vev,
we obtain the superpotential in the vacua $|k\rangle$ or $|k\rangle '$
for which all the fields have been integrated out,
\begin{equation}
\label{wlow}
\wl^{|k\rangle} = {N m^{3}\over 12 g^{2}} \Bigl( 1- \left(1-\la e^{2i\pi 
k/N}\right)^{3/2}\Bigr)\, ,\,\,
\wl^{|k\rangle '} = {N m^{3}\over 12 g^{2}} \Bigl( 1 + \left(1-\la e^{2i\pi 
k/N}\right)^{3/2}\Bigr)\, .
\end{equation}
One can also calculate the expectation values of $\tr\phi^{2}$ and 
$\tr\phi^{3}$,
\begin{equation}
\label{uvev}
\langle \tr\phi^{2}\rangle =2\, {\partial\wl\over\partial m}\,\cvp\quad
\langle\tr\phi^{3}\rangle = 3\,{\partial\wl\over\partial g}\,\cdotp
\end{equation}
\begin{figure}
\centerline{\epsfig{file=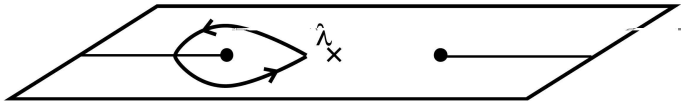,width=9cm}}
\caption{A closed path in the $\la$-plane for the ${\rm U}(2)$ theory. The 
bullets are at $\la =\pm 1$.
\label{fig0}}
\end{figure}

Formulas (\ref{Szvev}) and (\ref{wlow}) are singular (non-analytic) for the
special critical values (\ref{critlam}) of $\la$. This has a very
interesting consequence. Consider a closed path in the $\la$-plane, starting
in the vicinity of the weakly coupled point $\la =0$, and going through one
of the cut associated with the square roots in (\ref{Szvev}) or
(\ref{wlow}). We have depicted an example for ${\rm U}(2)$ in Figure 1.
Because the square root picks a minus sign in this process, the vacua
$|1\rangle$ and $|1\rangle'$ are continuously deformed into one another
when going along the path. More generally, the sheets of parameter space
corresponding to the vacua $|k\rangle$ and $|k\rangle'$ are connected
through branch cuts originating at the singular points $\la=\lck$. The
structure of the parameter space is thus drastically changed by quantum
effects. We will further discuss this kind of phenomenon in Section 3.

The model also has an interesting dependence in the bare $\theta$ angle. 
The transformation $\theta\rightarrow\theta + 2\pi$, or equivalently 
$\Lu^{2}\rightarrow\Lu^{2}e^{2i\pi /N}$, amounts to a 
non-trivial monodromy among the vacua. Since the different vacua are 
physically inequivalent, this means that the physical correlation 
functions of the theory, which satisfy the cluster decomposition principle, 
are not $2\pi$ periodic in $\theta$. The same phenomenon was discussed in 
detail on a two dimensional model in \cite{fertheta}.

\subsection{Massless glueball}

It is natural to guess that the non-analyticity at the critical values
(\ref{critlam}) is due to new degrees of freedom becoming massless.
This is easily demonstrated in our case. If
\begin{equation}
\label{epsdef}
\epsilon_{k} = 1-\la/\lck
\end{equation}
denotes the deviation from the singular point,
it is straightforward to show, by
using (\ref{wS}), (\ref{wp}) and (\ref{deltaeq}),
that the glueball superpotential
takes the following form for small $\epsilon_{k}$,
\begin{equation}
\label{wScrit}
W(S) = {2N\La^{3}\over 3\lck} - N\epsilon_{k} S + {N\lck^{2}S^{3}\over 
3\La^{6}} + \cdots\, \cvp
\end{equation}
where we have written only the most relevant terms. For $\epsilon_{k}
=0$, $W(S)$ is critical. This implies, under a mild assumption of
regularity for the K\" ahler potential, that we have a massless
glueball. We could have used the superpotential (\ref{wz}) for $z$ as
well. By introducing $\tilde z = z + m/(2g)$, we have
\begin{equation}
\label{wzcrit}
\ww^{(k)} = {Nm^{3}\over 12g^{2}} -
{Nm^{2}\epsilon_{k}\tilde z \over 4g}+ {gN\tilde z^{3}\over 3}\, \cdotp
\end{equation}
In the particular case we are considering, 
the superpotentials (\ref{wScrit}) and (\ref{wzcrit}) give equally 
valid descriptions of the low energy physics. The variables $S$ and 
$z$ are on the same footing because the derivative 
$\partial_{z}\partial_{S}\weff$ of (\ref{exW}) is non-zero at the 
critical point, and thus both $S$ and $z$ mix with the 
massless degrees of freedom. As we will explain in Section 4.4, for 
more general critical points the fields $S$ and $z$ do not 
necessarily play symmetric r\^oles.

\subsection{Confinement without a mass gap}

The appearance of massless degrees of freedom in the confining regime of
an ${\cal N}=1$ gauge theory is an interesting phenomenon. Note that
the critical points found in the preceding subsection are such that
the glueball condensate (\ref{Szvev}) vanishes,
\begin{equation}
\label{S0}
\langle S\rangle_{\rm crit} =0\, .
\end{equation}
If one considers, as was suggested in \cite{IS}, that $\langle
S\rangle$ is a good order parameter for confinement, equation
(\ref{S0}) would imply that the theory no longer confines at the critical
points. However, we are going to show that this interpretation is not
correct in general. Indeed, we can calculate explicitly the monopole
condensates at criticality, and we find
\begin{equation}
\label{monopolec}
\langle M_{m}\tilde M_{m}\rangle_{k} = i\sqrt{2}g\Lu^{2}e^{2i\pi k/N}
\sin {2\pi m\over N}\,\cvp\quad 1\leq m\leq N-1\, ,\quad {\rm for}\
\la=\lck\, .
\end{equation}
This demonstrates that for $N$ odd we have a non zero string tension 
in all magnetic ${\rm U}(1)$ factors, and thus confinement without a 
mass gap. For $N$ even the condensate for $m=N/2$ vanishes, and thus 
the corresponding electric charges do not confine (in particular 
for ${\rm U}(2)$ we don't have confinement at all). Those facts will be 
fully understood in Section 3. 

To derive (\ref{monopolec}), we start from the equation obtained by 
extremizing the superpotential (\ref{wquantum}) with respect to the 
$u_{p}$s and the monopole fields.
It is convenient to use the variables $x_{s}$ defined by
\begin{equation}
\label{xkdef}
u_{p}={1\over p} \tr\Phi^{p} = {1\over p}\sum_{s=1}^{N} x_{s}^{p}\, .
\end{equation}
We get
\begin{eqnarray}
\sqrt{2} \sum_{m=1}^{N-1} {\partial A_{D,m}\over\partial x_{s}}\, \langle 
M_{m}\tilde M_{m}\rangle &\!=\!& -m x_{s} - g x_{s}^{2}\, ,\label{meq}\\
A_{D,m} &\! =\! & 0\, .\label{meq2}
\end{eqnarray}
As explained in \cite{SW}, the variables $A_{D,m}$ are expressed
in terms of period integrals
\begin{equation}
\label{ADdef}
A_{D,m} = \oint_{\alpha_{m}} {x\, \d P\over 2i\pi y}
\end{equation}
over the hyperelliptic curve
\begin{equation}
\label{SWc}
y^{2} = P(x)^{2} - 4\Lu^{2N} = \prod_{s=1}^{N} (x-x_{s})^{2} - 
4\Lu^{2N}\, .
\end{equation}
The contours $\alpha_{m}$ encircle the cuts that vanish when the 
condition (\ref{meq2}) is satisfied. 
The calculation of $\partial A_{D,m}/\partial x_{s}$ was done in 
\cite{DS} in the $\suN$ theory. The $\uN$ theory we are dealing with 
presently is only slightly more general. It is possible to use the 
$\suN$ calculation by shifting the variables 
\begin{equation}
\label{tilddef}
x = \tilde x + z\, ,\quad x_{s} = \tilde x_{s} + z\, ,
\end{equation}
where $z$ is defined in (\ref{deffield}).
The constraints (\ref{meq2}) are solved in the $k^{\rm th}$ vacuum
by adjusting
\begin{equation}
\label{xkad}
\tilde x_{s} = 2\Lu e^{i\pi k/N}\cos {\pi (s-1/2)\over N}\,\cvp
\end{equation}
which corresponds to $P = 2(-1)^{k}\Lu^{N}\cos (N t)$ for $\tilde x =
2\Lu e^{i\pi k/N}\cos t$ \cite{DS}. It is then straightforward to show
that
\begin{equation}
\label{u1}
{\partial A_{D,m}\over\partial z} = \oint_{\alpha_{m}} {\d P\over 
2i\pi y} = \pm\oint_{\alpha_{m}}{N\,\d t\over 2\pi} = 0\, .
\end{equation}
The calculation of \cite{DS}, Section 2, can then be repeated without 
change, yielding
\begin{equation}
\label{solDS}
{\partial A_{D,m}\over\partial x_{s}} = {i\sin\hat t_{m}\over 
N(\cos t_{s} -\cos\hat t_{m})}
\end{equation}
for
\begin{equation}
\label{tdef}
t_{s} = {\pi (s-1/2)\over N}\,\cvp\quad 1\leq s\leq N\, ,\quad {\rm and}
\quad \hat t_{m} = {\pi m\over N}\,\cvp\quad 1\leq m\leq N-1\, .
\end{equation}
The right hand side of (\ref{meq}) can be evaluated by using 
(\ref{tilddef}), (\ref{xkad}) and (\ref{Szvev}),
\begin{equation}
\label{rhs}
-mx_{s}-gx_{s}^{2} = -2g\Lu^{2}e^{2i\pi k/N} \cos (2t_{s})\, .
\end{equation}
The trigonometric identity derived in \cite{DS},
\begin{equation}
\label{trigid}
\sum_{m=1}^{N-1} {\sin\hat t_{m}\sin (q \hat t_{m})\over 
N(\cos t_{s} -\cos\hat t_{m} )} = \cos (q t_{s})\, ,\quad q\in 
{\mathbb Z}\, ,
\end{equation}
together with (\ref{meq}), (\ref{solDS}) and (\ref{rhs}), yields 
(\ref{monopolec}).

\section{The quantum space of parameters}
\setcounter{equation}{0}

So far, we have adopted a purely field theoretical point of view. It is
possible to gain further insights by studying the geometrical
interpretation of our results, and in particular of the singular points on
the space of parameters. A geometrical interpretation is possible because
our model can be constructed in string theory by wrapping D5 branes on
special two-cycles of a certain non-compact Calabi-Yau manifold \cite{CIV}.
At least as far as the calculation of F-terms is concerned, this geometry
can be replaced by a dual geometry where the two-cycles go to three-cycles
and Ramond-Ramond flux through those cycles takes the place of the D5
branes \cite{CIV}. In theories with eight or more supercharges, the
singularities are usually associated with the degeneration of some cycle in
the geometry. Presently we are looking at a case with only four
supercharges for which very little is known. For the theory (\ref{lag}),
the non-trivial
part of the CY geometry is a simple non-compact hyperelliptic complex curve
given by an equation of the form
\begin{equation}
\label{cceq}
Y^{2} = \wt'(x)^{2} - R(x)\, ,
\end{equation}
where $R(x)$ is a polynomial of degree $p-2$ if the tree-level 
superpotential $\wt$ is a polynomial of degree $p$. The three-cycles 
of the original CY space correspond to one-cycles encircling the 
branch cuts of the surface (\ref{cceq}), and the non-zero RR fluxes 
are associated with non-zero period integrals of the differential 
form $Y\d x$. The $p-1$ free parameters in the polynomial $R(x)$ are 
then directly related to the fluxes through the $p-1$ cuts. A very 
useful way to look at the geometry (\ref{cceq}) is to realize that it 
comes from the solution of the one matrix model with potential given by 
$\wt$ \cite{DV}. The distribution of flux through the cycles depends 
on the pattern of gauge symmetry breaking. The solution of the matrix 
model with $C$ non-trivial cuts is associated with an unbroken gauge 
group of the form ${\rm U}(N_{1})\times\cdots\times {\rm U}(N_{C})$. 
We have collected in the Appendix a 
set of useful results on the multi-cut matrix models that are relevant 
to this problem. In the main text we limit the discussion to the cubic 
$\wt$ (\ref{Wtree}).

\subsection{One-cut solution}

Let us start with the one-cut solution that corresponds to the vacua
with unbroken gauge group that we have studied so far.
The RR flux goes
through one cycle only, and the other cycle is always degenerate. 
This means that the curve (\ref{cceq}) takes the special form
\begin{equation}
\label{cu1c}
Y^{2} = x^{2}(m+gx)^{2}-R(x)= M(x)^{2} (x-a)(x-b)\, .
\end{equation}
The non-trivial period integral is 
(for conventions see the Appendix and in 
particular Figure 4; units are chosen such that $\Lu=1$)
\begin{equation}
\label{f1}
\oint_{\gamma} Y\, \d x = -4i\pi S\, .
\end{equation}
It is easy to show that 
\begin{equation}
R(x) = 4Sg x + {1\over 2} \delta (S/\delta + m)(3S/\delta + m)
\end{equation}
where
\begin{equation}
\label{deltadef}
\delta = (b-a)^{2}/16
\end{equation}
is constrained by the condition (\ref{deltaeq}). Unlike cases with
${\cal N}=2$ supersymmetry, the most general geometry consistent with
the symmetries is not physical. The moduli are frozen by the condition
that the effective superpotential is extremal. As derived in
\cite{fer} (see also equations (\ref{t11}) and (\ref{pc1cut})), and
consistently with the equation (\ref{wp}), this amounts to imposing an
extremely simple condition,
\begin{equation}
\label{vac1}
\delta^{N} = 1 = \left({b-a\over 4}\right)^{2N}\, .
\end{equation}
The fact that this condition depends on $\wt$ only through the position of
the branch cuts $a$ and $b$ is a manifestation of the well-known
universality of matrix models. As explained in the Appendix, this property
remains true for any $\wt$ and an arbitrary number of cuts. Equation
(\ref{vac1}) shows that the non-trivial cycle of the physical curve can
never vanish. This implies that the singular points (\ref{critlam}), even
though they are associated with a vanishing period integral (\ref{f1}) as
stressed in (\ref{S0}), do not correspond to a vanishing cycle
$\gamma$.

One may wonder whether the singular points correspond to the old
Kazakov critical points of the matrix model \cite{kaz,MM}. The Kazakov
critical points occur when the root $x_{*}$ of the polynomial $M$ in
(\ref{cu1c}) coincides with one of the branching points $a$ or $b$. By
using (\ref{vac1}), is it immediate to see that
\begin{equation}
\label{xstar}
x_{*}= -{m\over g} - {a+b\over 2} =
-{m\over g}-\langle z\rangle\, ,
\end{equation}
where $\langle z\rangle = (a+b)/2$ is one of the $z$-field expectation
values (\ref{Szvev}). The equation $x_{*}=a$ or $x_{*}=b$ is then
solved when $\la$ goes to any of the $N$ values
\begin{equation}
\label{lKaz}
\la_{{\rm Kazakov},k}= {2\over 3}\, e^{-2i\pi k/N}\, ,\quad 0\leq k\leq 
N-1\, .
\end{equation}
Those are special values from the point of view of the matrix model,
and thus also from the point of view of the superpotential $W(S)$, 
but are clearly different from the critical values (\ref{critlam}). 
This comes from the fact that 
in gauge theory, only the solutions to $W'(S)=0$ have a 
physical significance. This is very different from the ordinary 
matrix model where the condition $W'=0$ has no meaning. In other words,
the gauge theory critical points are obtained when 
the solutions to $W'(S)=0$ are singular, and this condition is not 
related to the Kazakov condition of having $W$ singular.

It is not difficult to check that the genuine critical points 
(\ref{critlam}) occur when the double point of (\ref{cu1c})
sits exactly in the middle of the cut,
\begin{equation}
\label{middle}
x_{*} = {a + b\over 2}\,\cdotp
\end{equation}
This is a curious geometric condition that we will better understand in the 
next subsection.

\subsection{Two-cut solution}

\subsubsection{General properties}

The phases with broken gauge group $\uN\rightarrow {\rm
U}(N_{1})\times {\rm U}(N_{2})$, $N_{1}\not = 0$ and $N_{2}\not = 0$,
are described by the two-cut solution
of the matrix model. A full discussion and the derivations of some 
technical results used in the following can be found in the Appendix.
The curve (\ref{cceq}) now takes the form
\begin{equation}
\label{curve2c}
Y^{2}=g^{2} (x-a_{1})(x-b_{1})(x-a_{2})(x-b_{2})\, .
\end{equation}
The branch cuts are chosen to run from $a_{1}$ to $b_{1}$ and from 
$a_{2}$ to $b_{2}$.
The effective superpotential $W(S_{1},S_{2})$ can be calculated 
as a function of the fluxes through the
two cuts. Upon extremization, we obtain two conditions (\ref{Wextr}) 
that the curve (\ref{curve2c}) must satisfy. Those conditions
have $N_{1}N_{2}$ solutions corresponding to the vacua
$|k_{1},k_{2};N_{1},N_{2}\rangle$, $0\leq k_{j}\leq N_{j}-1$, of the
low energy ${\rm U}(N_{1})\times {\rm U}(N_{2})$ theory. 

Interesting points on parameter space are those for which the curve
(\ref{curve2c}) is singular. In the Appendix it is shown that the
vanishing of an electric cycle (which occurs if one of the cuts shrinks
to zero) is inconsistent with the conditions (\ref{Wextr}) except if
$N_{1}$ or $N_{2}$ is zero, while the vanishing of the magnetic cycle
(which occurs when the two cuts join) is possible if
\begin{equation}
\label{magc}
k_{1} \equiv k_{2}\ {\rm mod}\ N_{1}\wedge N_{2}\, , 
\end{equation}
where $N_{1}\wedge N_{2}$ is the greatest common divisor of $N_{1}$
and $N_{2}$. In particular, if $N_{1}$ and
$N_{2}$ are relatively prime, then all the vacua can have massless
monopole points. The vanishing of the magnetic cycle implies that the
electric coupling of the relative ${\rm U}(1)$ factor of the groups
${\rm U}(N_{1})$ and ${\rm U}(N_{2})$ blows up, or equivalently that the
magnetic coupling vanishes. We then have a massless magnetically
charged particle. Let us note that in both ${\cal N}=2$ and
${\cal N}=1$ gauge theories, a vanishing magnetic cycle is associated
with a massless magnetically charged particle. However, when the cycle
is non-vanishing, the mass of the monopole is exactly known only for
${\cal N}=2$ supersymmetry.

\begin{figure}
\centerline{\epsfig{file=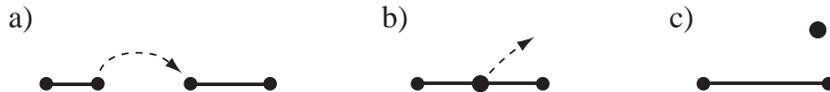,width=11cm}}
\caption{The transition from the two cut solution to the one cut solution 
(the inverse transition is of course possible as well). 
a) The endpoints of the two cuts get closer and 
eventually collide at the critical point, forming a double point. b) It is 
possible for the double point to move away from the cut. c) We are then 
left with the one-cut solution.
\label{figg}}
\end{figure}

Another simple but important result proved in the
Appendix is the following. Suppose that the parameters are adjusted in
such a way that a curve extremizing the superpotential
$W(S_{1},S_{2})$ has a vanishing magnetic cycle. This means that the
two cuts have joined to form a single cut.
{\it Then the curve is also a solution of the
extremization problem for the superpotential $W(S)$ relevant to the
one-cut solution.} Physically, this means that the ${\rm 
U}(N_{1})\times {\rm U}(N_{2})$ and $\uN$ branches
touch at the massless monopole point. One can 
go from the broken to the unbroken phase by condensing the massless 
monopole and the mass gap is created by the usual magnetic Higgs 
mechanism. The geometry of the transition is depicted in Figure 2.

\subsubsection{Examples}

The above discussion implies that in addition to the massless glueball
points discussed in Section 2, there will be many other singular
points with massless monopoles on the branches of parameter space
associated with the $\uN$ vacua $|k\rangle$ and $|k\rangle'$. Those
singularities are points of contact with the broken phases. The
calculation of the monopole condensates in Section 2.3 actually
suggests that for $N$ even the critical point (\ref{critlam}) is such
a point of contact, since we have found that the condensate $\langle
M_{N/2}\tilde M_{N/2}\rangle_{k}$ vanishes. We would then have a
massless monopole if $N$ is even in addition to the massless glueball
that is present for any $N$.

It is actually not difficult to check this picture explicitly, and to
discover that the relevant phase at $\la=\lc$ has ${\rm U}(N/2)\times
{\rm U}(N/2)$ unbroken. To do so, let us note that there are $N^{2}/4$
vacua $|k_{1},k_{2};N/2,N/2\rangle$ for the ${\rm U}(N/2)\times {\rm
U}(N/2)$ phase. The condition (\ref{magc}) is satisfied only in the
vacua for which $k_{1}=k_{2}$. This implies that the curve
(\ref{curve2c}) has a nice symmetry property that allows to solve the
constraint (\ref{Wextr}) directly. The solution, derived in the
Appendix, is
\begin{equation}
\label{Swbb}
Y^{2}=x^{2}(m+gx)^{2} - 4g^{2}e^{-4i\pi k/N}\Lu^{4}\, ,\quad 0\leq 
k_{1}=k_{2}=k\leq N/2 -1\, .
\end{equation}
This solution can also be found by using the results of
\cite{CIV} and \cite{CV}. As stressed in \cite{CV}, in the case $N=2$ 
the curve (\ref{Swbb}) exactly coincides with the Seiberg-Witten curve 
for ${\cal N}=2$ super Yang-Mills and gauge group ${\rm U}(2)$, with 
the moduli frozen to the classical values imposed by the tree-level 
superpotential. This means that there is an isomorphism in this 
particular case between the moduli space of ${\cal N}=2$ \cite{SW} 
and the parameter space of our ${\cal N}=1$ theory in the ${\rm 
U}(1)\times {\rm U}(1)$ phase. The monopole singularities are then 
nothing but the famous Seiberg-Witten singularities. For general even 
$N$, (\ref{Swbb}) degenerates precisely
at the $N$ critical points (\ref{critlam}), as we wished to prove.
The branches for $|k\rangle$ and $|k\rangle'$ touch the 
branch for $|k,k;N/2,N/2\rangle$ at $\la = \exp (-2i\pi k/N)$, and 
the branches for $|k+N/2\rangle$ and $|k+N/2\rangle'$ touch the 
branch for $|k,k;N/2,N/2\rangle$ at $\la = -\exp (-2i\pi k/N)$.

It should be clear that the massless monopole singularities are not in
general coinciding with (\ref{critlam}). To illustrate this point, let
us consider the ${\rm U}(3)$ theory. Since $N$ is odd we don't expect
to have a massless monopole for $\la=\lck$, but on the other hand the
condition (\ref{magc}) is satisfied for the broken ${\rm U}(2)\times
{\rm U}(1)$ phase. It is easy to see what happens explicitly, because the
solution can be found straightforwardly, for example by using the
results of \cite{CIV} and \cite{CV}. The curves for the vacua
$|\eta,0;2,1\rangle$, $\eta=\pm 1$, where classically two eigenvalues
are at zero and one is at $-m/g$, are given by
\begin{equation}
\label{solU3}
Y^{2} = (m+gx)(g x^{3} + mx^{2} + 4\eta g\Lu^{3})\, .
\end{equation}
A similar formula is valid for the vacua $|0,\eta;1,2\rangle$. The 
magnetic cycle of (\ref{solU3}) vanishes when $\la$ goes to
\begin{equation}
\label{crU3}
\la_{{\rm U}(2)\times {\rm U}(1),k}
= {8\over 9}\, e^{-2i\pi k/3}\, ,\quad 0\leq k\leq 2\, .
\end{equation}
It is then straightforward to check that at criticality
\begin{equation}
\label{critU3ab}
\left({b_{2}-a_{1}\over 4}\right)^{2} =\delta = e^{2i\pi k/3}\, .
\end{equation}
This is the correct condition for the one-cut phase. More precisely,
we have a connection with the ${\rm U}(3)$
vacua $|k\rangle$ and $|k\rangle'$ for $\la=\la_{{\rm U}(2)\times {\rm 
U}(1),k}$.

\begin{figure}
\centerline{\epsfig{file=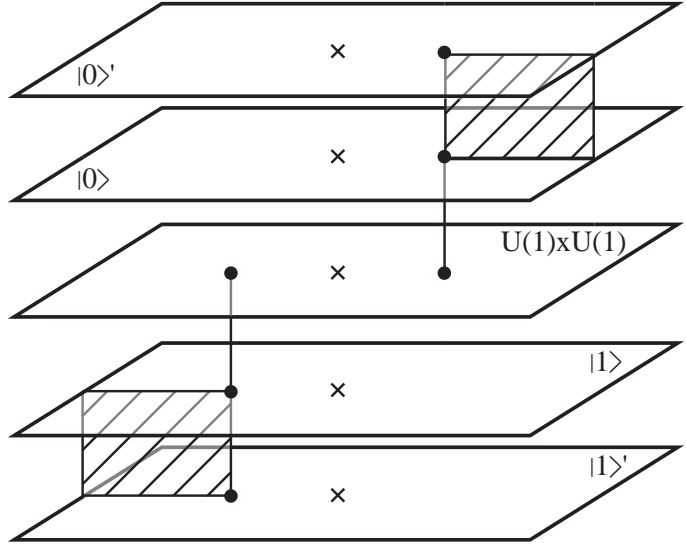,width=9cm}}
\caption{The five-sheeted structure of the parameter space for the 
${\rm U}(2)$ gauge theory. The crosses denote the $\la=0$ 
point, and bullets represent singularities with 
massless glueball and monopole.
\label{fig}}
\end{figure}
\subsection{The quantum parameter space for ${\rm U}(2)$}

As an illustration, we have depicted the full quantum space of 
parameters for the gauge group ${\rm U}(2)$ in Figure 
3. The quantum parameter space is connected, whereas its classical 
counterpart would have three disconnected components by excluding the point 
$\la=\infty$. A similar picture would 
be valid for ${\rm U}(N)$ with $N$ even, the ${\rm U}(1)\times {\rm 
U(1)}$ phase being replaced by ${\rm U}(N/2)\times {\rm U}(N/2)$ in the 
$|k,k;N/2,N/2\rangle$ vacua, and $|0\rangle$, $|0\rangle'$, $|1\rangle$, 
$|1\rangle'$ being replaced by $|k\rangle$, $|k\rangle'$, $|k+N/2\rangle$ 
and $|k+N/2\rangle'$. There are also other components with various 
inter-connections in that case.

\section{The large $N$ limit}
\setcounter{equation}{0}
\subsection{Generalities}

We can now tackle the problem that was the original motivation for
this work. We have found non-trivial critical points for some
particular values of the Higgs couplings. Following \cite{F1,F2}, we
would expect a non-trivial behaviour of the large $N$ expansion at the
critical points. The only observables that we can calculate exactly are
all related to the exact superpotentials discussed previously, or to
the electric ${\rm U}(1)$ coupling $\tau$ in the broken phases. To be
concrete, let us focus for the moment on the unbroken vacua and
discuss the tensions of domain walls. The coupling $\tau$ will be
discussed in section 4.3. The tension of a $(\p,\q)$ BPS domain wall
interpolating between vacua $\p$ and $\q$ is simply given by
\cite{DShif}
\begin{equation}
\label{T}
T_{\p,\q} =| \tau_{\p,\q} |
\end{equation}
where the complexified tension $\tau_{\p,\q}$ is defined by
\begin{equation}
\label{taudef}
\tau_{\p,\q} = N \left(\strut\smash{\wl^{\p} - \wl^{\q}}\right)\, .
\end{equation}
The factor of $N$ in (\ref{taudef}) comes from the normalization of
the $F$-term in (\ref{lag}). The standard lore about those domain
walls is based on the analysis of the $g=0$ theory \cite{Wbrane}. The
tension is given in that case by
\begin{equation}
\label{Tg0}
T_{\p,\q} = 2 N^{2} |\La|^{3} \sin {\pi |p-q|\over N}\,\cdotp
\end{equation}
The basic domain walls for which $|p-q|=1$, or more generally for which 
$|p-q|$ is of order one at large $N$, have a tension that scales as $N$ 
when $N\rightarrow\infty$,
\begin{equation}
\label{Ninf}
T_{\p,\q} \mathrel{\mathop{\kern 0pt\sim}\limits_{N\rightarrow\infty}^{}}
2\pi N |p-q| |\La|^{3}\, .
\end{equation}
This is consistent with a D-brane interpretation for the walls,
with a closed string coupling constant of order $1/N$. It was indeed
argued in \cite{Wbrane} that the confining strings can end on the
domain walls. More generally, with an arbitrary tree-level
superpotential, the formula (\ref{Ninf}) is replaced by
\begin{equation}
\label{Ninfg}
\tau_{\p,\q} \mathrel{\mathop{\kern 0pt\sim}\limits_{N\rightarrow\infty}^{}}
2i\pi (p-q) \Lu^{2}\partial_{\Lu^{2}}\wl = 2i\pi N (p-q) \langle 
S\rangle\, ,
\end{equation}
where the expectation value is taken in any vacuum $|k\rangle$ for which 
$|k-p|$ is of order one (we will take $k=0$). The standard D-brane 
interpretation is thus valid as long as $\langle S\rangle\not =0$.

Points for which $\langle S\rangle =0$ are special from the point of view 
of large $N$, but are not necessarily associated with a breakdown of the 
$1/N$ expansion. For any even tree-level superpotential, it is actually
very easy to adjust the parameters to get $\langle S\rangle =0$.
For example, with 
\begin{equation}
\label{wtex}
\wt (\Phi) = {m\over 2}\, \Phi^{2} + {g_{4}\over 4}\, \Phi^{4}\, ,
\end{equation}
we have \cite{fer}, in the vacua corresponding to 
$\langle\phi\rangle_{\rm cl}=0$,
\begin{equation}
\label{Svevquar}
\langle S\rangle = m\Lu^{2}e^{2i\pi k/N} + 3 g_{4}\Lu^{4}e^{4i\pi k/N}\, ,
\end{equation}
which is zero for
\begin{equation}
\label{sp}
m = -3g_{4}\Lu^{2}e^{2i\pi k/N}\, .
\end{equation}
The large $N$ expansion of the domain walls tensions is nevertheless
perfectly well-behaved, starting at order $N^{0}$ for the special
value (\ref{sp}). It is even very easy to get exactly tensionless $(\p,\q)$
domain walls, $\wl^{\p}=\wl^{\q}$, at finite $N$. By using the results of
\cite{fer} it is straightforward to see that this happens for example
for the theory (\ref{wtex}) when
\begin{equation}
\label{tensionless}
2m\sin {\pi (p-q)\over N} = -3 g_{4}\Lu^{2} e^{i\pi (p+q)/N} \sin {2\pi 
(p-q)\over N}\,\cdotp
\end{equation}
The condition (\ref{sp}) is recovered from the above equation
in the large $N$ limit.

In our theory, in addition to the $(\p,\q)$ domain walls, we have
$(\p',\q')$ domain walls with similar properties. More interestingly, there
are also $(\p',\q)$ domain walls. Those exist at the semi-classical level,
unlike the $(\p,\q)$ walls that originate from chiral symmetry breaking.
Their complexified tensions are simply given by
\begin{equation}
\label{taudefp}
\tau_{\p',\q} = N \left(\strut\smash{\wl^{\p'} - \wl^{\q}}\right) .
\end{equation}
In the semi-classical regime this is well approximated by
\begin{equation}
\label{tppcl}
\tau_{\p',\q}^{\rm cl} = {N^{2} m^{3}\over 6g^{2}}\,\cvp
\end{equation}
and scales as $N^{2}$
at large $N$. The $(\p',\q)$ walls thus behave like closed string
solitons (as opposed to D-branes). Remarkably, the results of Section 2
imply that the closed string solitons $(\p',\q)$ and the D-branes $(\p,\q)$
or $(\p',\q')$ can be continuously deformed into one another by varying the
parameters (see the discussion associated with Figure 1). 
This is reminiscent of the monodromy between
magnetic monopoles and quarks in strongly coupled ${\cal N}=2$ gauge
theories, as described explicitly for example in \cite{ferbil}. Moreover,
the $(|k\rangle,|k\rangle')$ domain wall is exactly tensionless at the
critical value $\la=\lck$ (\ref{critlam}), as can be checked easily by using
(\ref{wlow}). The presence of a tensionless solitonic domain wall is an 
important feature of our critical point, and will be associated with a 
singular $1/N$ expansion.

\subsection{Large $N$ and critical points}

From (\ref{Ninfg}) and (\ref{Szvev}), we have
\begin{equation}
\label{Ninf2}
\tau_{\p,\q} \mathrel{\mathop{\kern 0pt\sim}\limits_{N\rightarrow\infty}^{}}
2i\pi N (p-q)\La^{3}\sqrt{1-\la} \, ,
\end{equation}
which generalizes (\ref{Ninf}) to arbitrary $\la$. The same formula up
to a global minus sign is valid for $\tau_{\p',\q'}$, and from
(\ref{wlow}) we can also deduce
\begin{equation}
\label{Ninf3}
\tau_{\p',\q} \mathrel{\mathop{\kern 0pt\sim}
\limits_{N\rightarrow\infty}^{}}
{N^{2}m^{3}\over 6g^{2}}\, (1-\la)^{3/2} \, .
\end{equation}
A well-behaved large $N$ expansion would then predict that
$\tau_{\p,\q}$ and $\tau_{\p',\q'}$ are of order $N^{0}$ at $\la =\lc
=1$,\footnote{Without loss of generality, we focus on the critical point
$\la=\lc=1$ in the following.} while $\tau_{\p',\q}$ would be of order $N$,
but this is not what happens. The exact formulas show that 
\begin{eqnarray}
&&\tau_{\p,\q} \mathrel{\mathop{\kern 0pt\sim}
\limits_{N\rightarrow\infty}^{}}
{2^{5/2}\pi^{3/2}e^{i\pi /4}\over 3} \La^{3} \sqrt{N}
\left( q^{3/2}-p^{3/2}\right)\, ,\label{tcrit}\\
&&\tau_{\p',\q} \mathrel{\mathop{\kern 0pt\sim}
\limits_{N\rightarrow\infty}^{}}
{2^{5/2}\pi^{3/2}e^{i\pi /4}\over 3} \La^{3} \sqrt{N}
\left( q^{3/2}+p^{3/2}\right)\, ,\label{tcritb}
\end{eqnarray}
at criticality. The common $\sqrt{N}$ dependence for the ``soliton'' 
and the ``D-brane'' is remarkable and
signals the breakdown of the 
$1/N$ expansion near $\la=\lc$. It is straightforward to compute 
the $1/N$ corrections to (\ref{Ninf2}) or (\ref{Ninf3}) for $\la\not =\lc$,
\begin{eqnarray}
&& \hskip -1.7cm \tau_{\p,\q} = 2i\pi N (p-q)\La^{3}\sqrt{1-\la} \Biggl[
1 + {i\pi (2-3\la) (p+q)\over 2(1-\la) N} + {\cal O}\left(\strut
1/(N(1-\la))^{2}\right) \Biggr] ,\label{Nexp}\\
&&\hskip -1.7cm
\tau_{\p,\q'} = {N^{2}m^{3}\over 6g^{2}}\, (1-\la)^{3/2} \Biggl[
1 - {3i\pi \la (p+q)\over 2(1-\la) N} + {\cal O}\left(\strut1/(N(1-\la))^{2}
\right)\Biggr]\, .\label{Nexp2}
\end{eqnarray}
The expansions (\ref{Nexp}) and (\ref{Nexp2}) are
singular at $\la=1$ as expected.

\subsection{The double scaling limit}

The formulas (\ref{Nexp}) and (\ref{Nexp2}) are extremely suggestive.
The divergences at $\la  =1$ are very specific, and can be compensated
for by taking $N\rightarrow\infty$ and $\la\rightarrow \lc =1$ in a
correlated way given in (\ref{dsca}). The rescaled tensions
\begin{equation}
\label{tsdef}
t_{\p,\q} = \sqrt{1-\la} \,\tau_{\p,\q}\, ,\quad
t_{\p',\q} = \sqrt{1-\la} \,\tau_{\p',\q}
\end{equation}
then go to finite universal limits
\begin{eqnarray}
&& t_{\p,\q}^{\rm scaled} = {2\La^{3}\over 3\kappa^{2}} \Bigl[ 
\left(\strut 1-2i\pi q\kappa \right)^{3/2} -
\left( \strut 1-2i\pi p\kappa \right)^{3/2} \Bigr]\, ,\label{tsca}\\
&& t_{\p',\q}^{\rm scaled} = {2\La^{3}\over 3\kappa^{2}} \Bigl[ 
\left(\strut 1-2i\pi q\kappa \right)^{3/2} +
\left( \strut 1-2i\pi p\kappa \right)^{3/2} \Bigr]\, .\label{tsca2}
\end{eqnarray}
In the scaling (\ref{dsca}), it is natural to conjecture that the original
gauge theory reduces to a four dimensional non-critical string theory, or,
equivalently, to a five dimensional critical string theory. Equations
(\ref{tsca}) and (\ref{tsca2}) are interpreted as giving the exact tensions
for BPS D2-branes and solitonic two-branes in this string theory. The
rescaling (\ref{tsdef}) corresponds to a renormalization in the world-sheet
theory. A detailed discussion of this conjecture can be found in
\cite{F3,F2D,Frevue}, and it will not be repeated here. As explained in the
introduction, the idea is simply to generalize the old matrix model
approach to non-critical strings \cite{MM,matrev}. Note that by going
through the branch cuts in equations (\ref{tsca}) or (\ref{tsca2}), we can
transform continuously a D-brane (whose tension goes like $1/\kappa$ at
weak coupling) into a soliton (whose tension goes like $1/\kappa^{2}$ at
weak coupling), and vice-versa.

We have emphasized in section 3.1 that the critical points (\ref{critlam})
of the gauge theory are not the same as the critical points of the one-cut
matrix model. On the other hand, for $N$ even, the critical points are also
seen in the two-cut matrix model, and there they do correspond to a regime
that was used to describe the $c=1$ strings \cite{c1str,KKK}. We would like
to stress that this fact is, as far as we can see, of no deep significance
in the present context. It does mean that large Feynman graphs dominate
near the critical points. This is perfectly consistent with the four
dimensional path integral picture sketched in the introduction because the
matrix model planar diagrams are related to gauge theory planar diagrams
\cite{DV}. However, the size $N$ of matrices in the gauge theory path
integral is not related to the size $n$ of matrices in the matrix model. In
particular, the scaling (\ref{dsca}) relevant to the gauge theory is not
the same as the scaling relevant to the $c=1$ matrix model that was worked
out long ago in \cite{c1sca} (four dimensional scalings reminiscent of the
$c=1$ scaling do occur \cite{F3}, but in different cases). The crucial
point is that the double scaling limit (\ref{dsca}) yields a four
dimensional non-critical (or five dimensional critical) string, because the 
starting point is a four dimensional path integral. This is very different
from the $c=1$ string. The gauge theory path integral being
non-perturbatively defined, the scaling provides a full non-perturbative
definition of the resulting string theory. Again this is in sharp contrast
with the $c=1$ string case.

By using (\ref{Swbb}), (\ref{for4}) and (\ref{kkdef}), it is easy to study
the scaling of the ${\rm U}(1)$ coupling $\tau_{p}$ of the
$|p,p;N/2,N/2\rangle$ vacua. It is actually convenient to work with the
dual magnetic coupling
\begin{equation}
\label{taudual}
\tau_{p,\rm D} = -1/\tau_{p}\, .
\end{equation}
The parameter $k'^{2}$
of the curve (\ref{Swbb}) goes to
\begin{equation}
\label{k2par}
k'^{2} \longrightarrow 2\sqrt{2}\sqrt{1-\la}\sqrt{1-2i\pi p\kappa}
\end{equation}
in the scaling (\ref{dsca}). The renormalized coupling
\begin{equation}
\label{Tren}
t_{p,\rm D} = \tau_{p,\rm D} - {i\over\pi}
\ln {\sqrt{1-\la}\over 4\sqrt{2}}
\end{equation}
then goes to a finite limit
\begin{equation}
\label{tausca}
t_{p,\rm D}^{\rm scaled} = {i\over 2\pi} \ln (1-2i\pi p \kappa)\, .
\end{equation}

There is a subtle difference between the double scaling limit yielding
(\ref{tsca}), (\ref{tsca2}) or (\ref{tausca}) and the double scaling
limits discussed in previous papers \cite{F3,F2D,Frevue}. Even though
equations (\ref{Ninf2}) or (\ref{Ninf3}) clearly shows that the
divergences encountered in $1/N$ corrections have an IR origin, the
world-sheet renormalizations (\ref{tsdef}), whose form are dictated by
the $N$ dependence of $\tau_{\p,\q}$ and $\tau_{\p',\q}$, correspond
to a UV limit in space-time.

\subsection{Multicritical points}

Let us sketch an elementary field theoretic discussion of more general
critical points. A systematic study can certainly be done by using the
ideas described in Section 3, but this is beyond the scope
of the present paper. Let us consider an arbitrary tree level 
superpotential of degree $p$,
\begin{equation}
\label{wtgen}
\wt (\Phi) = \sum_{r=1}^{p} {g_{r}\over r}\, \Phi^{r}\, .
\end{equation}
Classically, the theory has generically $p-1$ independent vacua with
unbroken gauge group, labeled by an integer $J$. Our goal is to construct 
multicritical points akin to the one studied in Section 2. 

A first important step is
to understand the distinction between the variable $z=(\tr\phi)/N$,
which is natural from the UV, ${\cal N}=2$ point of view, and the
glueball field $S$, which is more natural from the IR, ${\cal N}=1$
point of view.\footnote{I would like to thank N.~Seiberg for raising
this point.} There are $N$
superpotentials for $z$, each of degree $p$, and labeled by an
integer $k$. The formula generalizing (\ref{wz}) to the case of
(\ref{wtgen}) was derived in \cite{fer} and reads
\begin{equation}
\label{wzgen}
\ww^{(k)}(z) = N\sum_{r\geq 0} \omega_{r}^{(k)} z^{r}\, ,
\end{equation}
where
\begin{equation}
\label{omega}
\omega_{r}^{(k)} = \sum_{q\geq 0}{g_{r + 2 q}\over r+2q}
{\rm C}_{r+2q}^{2q} {\rm C}_{2q}^{q} \Lu^{2q} e^{2i\pi kq/N}\, .
\end{equation}
The equation $\ww^{(k)'}(z)=0$ has $p-1$ solutions, corresponding to
the $p-1$ classical vacua. It is impossible, by considering a given
superpotential for $z$ (or for any of the fields $\tr\phi^{r}$), to
derive the existence of $N$ vacua associated with each of the $p-1$
classical vacua. This is in sharp contrast with the superpotentials
for the field $S$. There are $p-1$ of them, that we denote
$W_{(J)}(S)$, $0\leq J\leq p-2$. Each of the equations $W_{(J)}'(S)=0$
has precisely $N$ solutions, reflecting chiral symmetry breaking in
the pure ${\cal N}=1$ theory.

Critical points of any order for the
field $z$ can be straightforwardly obtained. For example, formulas
(\ref{wzgen}) and (\ref{omega}) imply that we have a critical point of
order $\ell$ at $z=0$ when the $g_{r}$s are such that
$\omega_{r}^{(k)}=0$ for $1\leq r\leq\ell$. However, it is important
to realize that an $\ell^{\rm th}$ order critical point of
$\ww^{(k)}(z)$ does not necessarily correspond to an $\ell^{\rm th}$
order critical point for a corresponding glueball superpotential
$W_{(J)}(S)$. Let us give a concrete example based on the tree-level
superpotential (\ref{wtex}), which yields
\begin{equation}
\label{wkexqu}
\ww^{(k)}(z) =Nm \Lu^{2}e^{2i\pi k/N} + {3N\over 2}\, g_{4}\Lu^{4}e^{4i\pi 
k/N} + {N\over 2}\left( m + 6g_{4}\Lu^{2}e^{2i\pi 
k/N}\right) z^{2} + {Ng_{4} z^{4}\over 4}\,\cdotp
\end{equation}
There are $3N$ vacua with unbroken gauge group,
denoted by $|k\rangle^{J}$, $0\leq J\leq 2$, $0\leq k\leq N-1$, for 
which
\begin{eqnarray}
&&\hskip -.7cm \langle S\rangle_{|k\rangle^{0}} = m\Lu^{2}e^{2i\pi 
k/N} + 3g_{4}\Lu^{4}e^{4i\pi k/N} \, ,\quad
\langle z\rangle_{|k\rangle^{0}} = 0 \, ,\\
&&\hskip -.7cm\langle S\rangle_{|k\rangle^{1}} =-2m\Lu^{2}e^{2i\pi 
k/N} - 15g_{4}\Lu^{4}e^{4i\pi k/N} \, ,\quad\!\!
\langle z\rangle_{|k\rangle^{1}} = i\Bigl( {m\over 
g_{4}}\Bigr)^{1/2}\Bigl( 1+ {6g_{4}\Lu^{2}e^{2i\pi k/N}\over 
m}\Bigr)^{1/2}\! , \nonumber\\
&&\hskip -.7cm \langle S\rangle_{|k\rangle^{2}} =-2m\Lu^{2}e^{2i\pi 
k/N} - 15g_{4}\Lu^{4}e^{4i\pi k/N} \, ,\quad\!\!
\langle z\rangle_{|k\rangle^{2}} = -i\Bigl( {m\over 
g_{4}}\Bigr)^{1/2}\Bigl( 1+ {6g_{4}\Lu^{2}e^{2i\pi k/N}\over 
m}\Bigr)^{1/2}\!\!\! .\nonumber\label{Szvev22}
\end{eqnarray}
There is a
critical point for $\ww^{(k)}(z)$ when $m = -6g_{4}\Lu^{2}
e^{2i\pi k/N}$. The field $z$ is then massless in the vacua
$|k\rangle^{J}$, for any $J$. On the other hand, it is straightforward
to check, by using the results in \cite{fer},
that the superpotentials $W_{(J)}(S)$ are not critical, and thus we do 
not have a massless glueball.
The non-trivial phenomena described in the present
paper, like the breakdown of the large $N$ expansion, only occur in
the vacua with massless glueballs. For the other critical points,
the large $N$ expansion is perfectly well-behaved.

Multicritical points associated with a singular large $N$ expansion
can be obtained by considering
a general odd tree-level superpotential. When only $g_{1}$ 
and $g_{3}$ are turned on, the critical point at $z=0$ for 
$g_{1}=-2g_{3}\Lu^{2}$ is in the same universality class as
(\ref{critlam}). If we turn on $g_{5}$, we can go to a higher 
critical point for $g_{1}=18 \Lu^{4}e^{4i\pi k/N}g_{5}$ and $g_{3}=
-12\Lu^{2}e^{2i\pi k/N} g_{5}$. We then have $\ww^{(k)}(z) = 
Ng_{5}z^{5}/5$. We have checked explicitly using results 
in \cite{fer} that the superpotential for the glueball 
superfield also goes like $S^{5}$ at criticality. The tension of the 
domain walls then goes like $N^{3/4}$ at large $N$. The same 
construction starting with an odd $\wt$ of degree $2\ell +1$ 
presumably yields similar critical points with $\ww 
(z)\propto z^{2\ell +1}$ and $W(S)\propto S^{2\ell +1}$.
The large $N$ tension at criticality scales as
\begin{equation}
\label{tenc}
T(\ell) \propto N^{1-1/(2\ell)}\, ,
\end{equation}
and double scaling limits can certainly be defined. It would be nice 
to work out those multicritical points and the associated double 
scaling limits more explicitly.

\section{Conclusion and prospects}
\setcounter{equation}{0} 

Non-trivial exact ${\cal N}=1$ effective superpotentials 
have proven to be extremely powerful tools to work out some new interesting 
physics in strongly coupled ${\cal N}=1$ gauge theories. There is a 
qualitative similarity with ${\cal N}=2$ gauge theories, the parameter 
space replacing the moduli space. There are also some 
fundamental differences. For example, the singularities are not
necessarily associated with vanishing cycles in the geometric description, 
and extended objects play an important r\^ole. 
It would be nice to understand the general structure 
of the quantum space of parameters for an arbitrary polynomial $\wt$, 
and in particular to study higher critical points \`a la 
Argyres-Douglas.

The matrix model proposal made in \cite{DV} can in principle be used
to study a wide class of examples, and we are presently working on the 
theory with two adjoint Higgs fields. One of the motivations 
to study such a model is that it is not a simple deformation of a 
theory with extended supersymmetry, unlike all the cases that have been 
worked out for the moment \cite{DV,fer,DH}.

Maybe the most important message of this paper is that the old matrix model
approach to non-critical strings can be generalized to four dimensional
theories with ${\cal N}=1$ supersymmetry. This is a new and very important
example where the ideas advocated in \cite{F1,F2,F3,F2D,Fnp,Frevue} apply.
The results for ${\cal N}=2$ obtained in \cite{F3} actually apply in ${\cal
N}=1$ with a degree $N$ tree-level superpotential. The parameter space for
the phase with maximal gauge symmetry breaking $\uN\rightarrow {\rm
U}(1)^{N}$ is indeed isomorphic to the ${\cal N}=2$ moduli space \cite{CV}.
It seems that much could be learned on four dimensional non-critical
strings in this way. Only very few results are available, and we believe
that it will be extremely rewarding to work out the general structure
behind the four dimensional double scaling limits. The study of gauge
theories with adjoint Higgs fields in two or three dimensions, and the
associated critical points and double scaling limits, could also be
potentially very interesting.

\subsection*{Acknowledgements}

This work was initiated thank's to Robbert Dijkgraaf inspiring talk at
the Strings 2002 conference in Cambridge, UK. I am also indebted to
Edward Witten for suggesting that the classical vacua
$\langle\phi\rangle_{\rm cl} =0$ and $\langle\phi\rangle_{\rm cl} =
-m/g$ may mix at strong coupling. I would also like to acknowledge
many useful discussions with J.-P.~Derendinger, R.~Hern\'andez,
T.~Hollowood, K.~Intriligator, N.~Seiberg and V.~Kazakov. This work
was supported in part by the Swiss National Science Foundation.

\renewcommand{\thesection}{\Alph{section}}
\renewcommand{\thesubsection}{\arabic{subsection}}
\renewcommand{\theequation}{A.\arabic{equation}}
\setcounter{section}{0}
\section*{Appendix: The multi-cut solutions}
\setcounter{equation}{0}
\subsection{Generalities}

Let us imagine that we are considering the theory (\ref{lag}) with an 
arbitrary polynomial tree-level superpotential of degree $p$
\begin{equation}
\label{wttgen}
\wt (\phi) = \sum_{r=1}^{p} {g_{r}\over r}\, \phi^{r}\, .
\end{equation}
The most general gauge symmetry breaking pattern is of the form
$\uN\rightarrow {\rm U}(N_{1})\times \cdots\times {\rm U}(N_{C})$, with
$N_{1}+\cdots + N_{C} = N$ and $C\leq p-1$. The quantum effective
superpotential is expressed in such a vacuum in terms of a
$N_{k}$-independent prepotential ${\cal F}$ as
\begin{equation}
\label{weffgen}
W = -\sum_{k=1}^{C}N_{k}\partial_{S_{k}}{\cal F}\, .
\end{equation}
The prepotential is given by the planar 
approximation to a holomorphic integral over complex $n\times n$ 
matrices \cite{DV},
\begin{equation}
\label{DVint}
\exp\left( n^{2}{\cal F} /S^{2}\right) = \int_{\rm planar}
\hskip -.5cm\d^{n^{2}}\phi\,
\exp\Bigl[ -{n\over S}\tr\wt(\phi)\Bigr]\, ,
\end{equation}
where we are working in units for which $\Lu=1$.
We can restrict ourselves to hermitian matrices and real couplings 
to compute (\ref{DVint}) because there is no ambiguity in the analytic 
continuation for planar diagrams.
The eigenvalue distribution $\rho (x)$ has a support
\begin{equation}
\label{supportdef}
{\rm Support}[\rho] = \bigcup_{k=1}^{C} [a_{k},b_{k}]
\end{equation}
on $C$ cuts $[a_{k},b_{k}]$ which classically shrink to points that
coincide with $C$ distinct roots of the equation $\wt'(x)=0$.
The prepotential
\begin{equation}
\label{Fform}
{\cal F} = -S \int\!\d x\, 
\rho(x)\wt (x) + S^{2}\int\!\d x\d z\, 
\rho(x)\rho(z)\ln |x-z|\, ,
\end{equation}
as well as the superpotential $W$, depend on the filling fractions
\begin{equation}
\label{ff}
{S_{k}\over S}= 
\int_{a_{k}}^{b_{k}}\rho(x)\, \d x
\end{equation}
that must be kept fixed in the integral (\ref{DVint}). The filling 
fractions satisfy the constraint
\begin{equation}
\label{sff}
\sum_{k=1}^{C}S_{k}=S\, .
\end{equation}
It is very convenient to introduce
\begin{equation}
\label{odef}
\omega (x) = \int_{-\infty}^{+\infty} {\rho (z)\, 
\d z\over x - z}\,\cvp
\end{equation}
in terms of which
\begin{equation}
\label{rvo}
\rho(x) = {i\over 2\pi} \left(\strut\omega (x +i\epsilon) - 
\omega (x -i\epsilon)\right)\, .
\end{equation}
The fundamental saddle point equation reads
\begin{equation}
\label{speq}
\wt '(x) = S\left(\strut \omega( x+i\epsilon) + \omega( 
x-i\epsilon)\right)\quad {\rm for}\quad x\in {\rm Support}\,[\rho]\, .
\end{equation}
The force acting on a test eigenvalue at $x$ 
is deduced from (\ref{Fform}) to be $-n Y(x)/S$ where
\begin{equation}
\label{Ydef}
Y(x) = \wt '(x) - 2S\omega (x)\, .
\end{equation}
One can show using (\ref{speq}) that $Y(x)$ satisfies an algebraic
equation \cite{matrev}
\begin{equation}
\label{yeq}
Y^{2} = \wt '(x)^{2} - R(x)=M(x)^{2}\prod_{k=1}^{C} (x-a_{k})(x-b_{k})\, ,
\end{equation}
where
\begin{equation}
\label{R}
R(x) = 4S g_{p} x^{p-2} + \cdots
\end{equation}
is a polynomial of degree $p-2$ and $M$ is a polynomial of degree 
$p-1-C$. The coefficients of $R$ are fixed in 
terms of the $S_{k}$
by the conditions (\ref{ff}) which can be conveniently rewritten
\begin{equation}
\label{cy1}
\oint_{\gamma_{k}}Y\,\d x = -4i\pi S_{k}\, .
\end{equation}
The definition of various contours is given in Figure 4.

\begin{figure}
\centerline{\epsfig{file=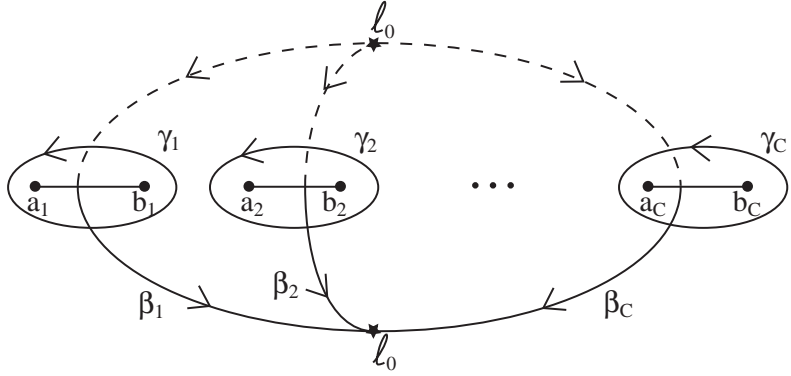,width=11cm}}
\caption{Definition of the contours $\gamma_{k}$ and $\beta_{k}$ used 
in the text (similar definitions were given in \cite{CV}).
The points $\ell_{0}$ and $\bar\ell_{0}$ are going to infinity on the 
upper and lower sheets of the non-compact Riemann surface (\ref{RS}). As a
consequence, the contours $\beta_{k}$ are not closed.
\label{fig3}}
\end{figure}

Since we are interested in the derivatives of $\cal F$ to compute the 
superpotential (\ref{weffgen}), it is very useful to introduce
\begin{equation}
\label{psidef}
\psi_{k}(x) = {\partial (S\omega)\over\partial S_{k}} = 
-{1\over 2} {\partial Y\over\partial S_{k}}\,\cdotp
\end{equation}
Equation (\ref{speq}) implies that
\begin{equation}
\label{ccpsi}
\psi_{k}(x+i\epsilon) + \psi_{k}(x-i\epsilon) = 0\quad {\rm for}\quad
x\in {\rm Support}\,[\rho]\, ,
\end{equation}
and (\ref{Ydef}) and (\ref{cy1}) imply that
\begin{equation}
\label{cnor}
{1\over 2i\pi} \oint_{\gamma_{l}}\psi_{k}(x)\,\d x = \delta_{k,l}\, .
\end{equation}
The asymptotics at infinity of $\psi_{k}$ are deduced from the
corresponding asymptotics $\omega (x)\sim 1/x$ and (\ref{psidef}),
\begin{equation}
\label{aspsi}
\psi_{k}(x) \mathrel{\mathop{\kern 0pt\sim}
\limits_{x\rightarrow\infty}^{}} {1\over x}\,\cdotp
\end{equation}
These results prove that the differentials $\psi_{k}\d x$ form a
canonical basis of $\log$-nor\-ma\-li\-za\-ble holomorphic one-forms on the
genus $C-1$ non-compact Riemann surface
\begin{equation}
\label{RS}
y^{2} =\prod_{k=1}^{C} (x-a_{k})(x-b_{k})\, .
\end{equation}
Note that the polynomial $M$ in (\ref{yeq}) no longer appears. This 
simplification is at the origin of ``universality'' in our problem: the 
derivatives of $W$ will depend on $\wt$ only through the branching points 
$a_{k}$ and $b_{k}$. We can write explicitly
\begin{equation}
\label{psiexpl}
\psi_{k}(x) = {N_{k}(x)\over y}\,\cvp
\end{equation}
where
\begin{equation}
\label{Nkex}
N_{k}(x) = x^{C-1} + \cdots
\end{equation}
is a polynomial of degree $C-1$ whose coefficients are determined by the
equations (\ref{cnor}). From (\ref{psiexpl}), (\ref{psidef}) and
(\ref{rvo}), one then obtains
\begin{equation}
\label{derrho}
{\partial (S\rho )\over\partial S_{k}} = \left\{ \matrix{
 {\displaystyle N_{k}(x)\over
\displaystyle\pi\sqrt{(x-a_{q})(b_{q}-x)\smash{\prod_{k\not = q}}
(x-a_{k})(x-b_{k})}}\quad & {\rm for}\ x\in [a_{q},b_{q}]\, ,\cr &\cr &\cr
0\quad &{\rm for}\ x\notin {\rm Support}[\rho]\, .\cr}\right.
\end{equation}
This equation can be used to compute the derivatives of $\cal F$, 
generalizing a calculation made in the Appendix of \cite{fer}. The first 
derivatives of $\cal F$ can also be obtained by noting that adding 
an eigenvalue to the $k^{\rm th}$ cut mimics the variation $\delta 
S_{k} = S/n$ \cite{DV}. By taking into account the energy cost in 
creating the eigenvalue at infinity, we then obtain
\begin{equation}
\label{dF1}
\partial_{S_{k}}{\cal F} = \lim_{\ell_{0}\rightarrow\infty}\Bigl(
{1\over 2}\int_{\beta_{k}} Y\,\d x + 2 S
\ln\ell_{0} - \wt (\ell_{0})\Bigr) \, ,
\end{equation}
where the contours $\beta_{k}$ are defined in Figure 4. By using 
(\ref{psidef}) we deduce
\begin{equation}
\label{dF2}
\partial_{S_{k}}\partial_{S_{l}}{\cal F} = -2i\pi t_{kl}
\end{equation}
or equivalently
\begin{equation}
\label{dW1}
\partial_{S_{k}}W = 2i\pi\sum_{l=1}^{C}N_{l}\, t_{kl}\, ,
\end{equation}
where $t_{kl}$ is the regularized ``period'' matrix of the non-compact
curve (\ref{RS}),
\begin{equation}
\label{periodm}
t_{kl} = \lim_{\ell_{0}\rightarrow\infty}\Bigl( {1\over 2i\pi}
\int_{\beta_{l}}\psi_{k}\,\d x + {i\over\pi} \ln\ell_{0}\Bigr)=t_{lk}\, .
\end{equation}
By using the normalizations (\ref{cnor}), it is straightforward to 
show that the trivial exchange of the endpoints of the $k^{\rm th}$ cut
amounts to the transformation $t_{kk}\rightarrow t_{kk} + 1$.
By taking into account this identification, we obtain that the
conditions for a critical point of $W$ are
\begin{equation}
\label{crgenW}
\sum_{l=1}^{C}N_{l}t_{jl} \equiv k_{j}\ {\rm mod}\
N_{j}\, ,\quad 1\leq j\leq C\, .
\end{equation}
There are $\prod_{j=1}^{C}N_{j}$ solutions to (\ref{crgenW}), labeled 
by the ${\mathbb Z}_{N_{j}}$-valued numbers $k_{j}$.
As already mentioned, an important property of (\ref{crgenW}) is that it 
depends on $\wt$ only through the positions of the branch cuts of the 
curve (\ref{RS}).

\subsection{Applications}

\subsubsection{One-cut solution}

In the one cut case, there are only two branching points $a_{1}=a$ and 
$b_{1}=b$, and the holomorphic one-form is
\begin{equation}
\label{hol1f1c}
\psi\,\d x = {\d x\over\sqrt{(x-a)(x-b)}}\,\cdotp
\end{equation}
The only period is 
\begin{equation}
\label{t11}
t_{11} = \lim_{\ell_{0}\rightarrow\infty}\Bigl(
{1\over i\pi}\int_{b}^{\ell_{0}} {\d x\over\sqrt{(x-a)(x-b)}}
+ {i\over\pi}\ln\ell_{0}\Bigr) = 
{i\over\pi}\ln {b-a\over 4}\,\cvp
\end{equation}
and the physical condition (\ref{crgenW}) is
\begin{equation}
\label{pc1cut}
N t_{11} \equiv k\ {\rm mod}\ N\, .
\end{equation}
We thus find again (\ref{vac1}).

\subsubsection{Two-cut solution}

We limit the discussion below to the two-cut solution
$\uN\rightarrow {\rm U}(N_{1})\times {\rm
U}(N_{2})$, $N_{1}\not = 0$ and $N_{2}\not = 0$, because this is the 
case used in the main text, but most of the arguments can be 
generalized straightforwardly to any number of cuts.
It is convenient to introduce the cycles $\alpha$ and $\gamma$ defined by
\begin{equation}
\label{magcy}
\alpha = \beta_{1}-\beta_{2}\, ,\quad\gamma = \gamma_{1}+\gamma_{2}\, .
\end{equation}
Simple identities are
\begin{equation}
\label{useid}
\oint_{\gamma} {\d x\over y} = 0\, ,\quad \oint_{\gamma} {x\, \d x\over 
y} = 2 i\pi\, .
\end{equation}
The basis of $\log$-normalizable holomorphic one-forms is
\begin{equation}
\label{basishol}
\psi_{1}\,\d x = {x + \mu_{1}\over y}\, \d x\, ,\quad
\psi_{2}\,\d x = {x + \mu_{2}\over y}\, \d x\, .
\end{equation}
By using (\ref{useid}) and $t_{12}=t_{21}$ one can derive the set of useful
formulas
\begin{eqnarray}
&&\mu_{2}-\mu_{1} = {2i\pi\over\oint_{\gamma_{2}}\d x/y}
 = {\pi\sqrt{(b_{2}-b_{1})(a_{2}-a_{1})}\over 2 K(k)}\, \cvp\label{for1}\\
&& \mu_{1}\oint_{\alpha}{\d x\over y} = 
(\mu_{1}-\mu_{2})\int_{\beta_{1}} {\d x\over y} - \oint_{\alpha} {x\, \d 
x\over y}\, \cvp\label{for2}\\
&& \mu_{2}\oint_{\alpha}{\d x\over y} = 
(\mu_{1}-\mu_{2})\int_{\beta_{2}} {\d x\over y} - \oint_{\alpha} {x\, \d 
x\over y}\, \cvp\label{for3}\\
&& 2t_{12}-t_{11}-t_{22} = \tau=
{\oint_{\alpha}\d x/y\over\oint_{\gamma_{2}}\d 
x/y} = {iK(k')\over K(k)}\,\cvp\label{for4}
\end{eqnarray}
where $K$ is the standard complete elliptic integral of the first kind and 
\begin{equation}
\label{kkdef}
k^{2} = {(b_{2}-a_{2})(b_{1}-a_{1})\over 
(b_{2}-b_{1})(a_{2}-a_{1})}\,\cvp\quad k'^{2} = 1-k^{2} =
{(a_{2}-b_{1})(b_{2}-a_{1})\over (b_{2}-b_{1})(a_{2}-a_{1})}\, \cdotp
\end{equation}
The parameter $\tau$ in (\ref{for4}) is the modular parameter of the
compact part of the curve (\ref{RS}). Physically it corresponds to the 
non-trivial electric coupling constant of the relative ${\rm U}(1)$
factor $(N_{2}\times {\rm U}(1)_{1} - N_{1}\times {\rm U}(1)_{2})/N$
of the ${\rm U}(N_{1})$ and ${\rm U}(N_{2})$ groups, where ${\rm 
U}(1)_{j}\subset {\rm U}(N_{j})$. 

Interesting physics is potentially associated with a singular curve, and it 
is important to understand how the periods of the smooth curve behave in 
the limit in which one of the cycle vanishes. Let us first consider the 
case of a vanishing electric cycle, for example $\gamma_{1}\rightarrow 0$. 
In the limit, we have $a_{1}=b_{1}=a$, and
\begin{equation}
\label{singl1}
\lim_{\gamma_{1}\rightarrow 0} \oint_{\gamma_{1}} {\d x\over y} =
{2i\pi\over\sqrt{(a_{2}-a)(b_{2}-a)}}\,\cvp\quad
\lim_{\gamma_{1}\rightarrow 0} \oint_{\gamma_{1}} {x\, \d x\over y} =
{2i\pi a\over\sqrt{(a_{2}-a)(b_{2}-a)}}\,\cdotp
\end{equation}
From those equations we deduce
\begin{equation}
\label{smu1}
\mu_{1} = -a + \sqrt{(a_{2}-a)(b_{2}-a)}\, ,\quad \mu_{2} = -a\, .
\end{equation}
One then shows immediately that the periods for the singular curve are
\begin{equation}
\label{tsing1}
t_{22} = {i\over \pi} \ln {b_{2}-a_{2}\over 4}\,\cvp\quad 
t_{12} = {i\over\pi}\ln { b_{2}-a_{2}\over 4}
{\sqrt{b_{2}-a} - \sqrt{a_{2}-a}
\over\sqrt{b_{2}-a} + \sqrt{a_{2}-a}}\,\cvp\quad
t_{11} = \infty\, .
\end{equation}
The divergence of $t_{11}$ is directly related to the vanishing of the 
electric coupling constant. Let us now 
consider the physically more interesting case of a vanishing magnetic 
cycle $\alpha\rightarrow 0$. We have $a_{2}=b_{1}=c$ in the limit, and 
we will then note $a_{1}=a$ and $b_{2}=b$. Formulas similar to 
(\ref{singl1}) are
\begin{equation}
\label{singl2}
\lim_{\alpha\rightarrow 0} \oint_{\alpha} {\d x\over y} =
{2\pi\over\sqrt{(c-a)(b-c)}}\,\cvp\quad
\lim_{\alpha\rightarrow 0} \oint_{\alpha} {x\, \d x\over y} =
{2\pi c\over\sqrt{(c-a)(b-c)}}\,\cdotp
\end{equation}
Moreover, $\oint_{\gamma_{2}}\d x /y$ diverges. This is equivalent to the
vanishing of $\tau$ or of the magnetic coupling. Equations
(\ref{for1}), (\ref{for2}) and (\ref{for3}) then imply
\begin{equation}
\label{sing5}
\mu_{1} = \mu_{2} = -c\, .
\end{equation}
For the singular curve we thus have 
\begin{equation}
\label{psising}
\psi_{1} = \psi_{2} = {1\over\sqrt{x-a)(x-b)}} = \psi\, ,
\end{equation}
and we immediately deduce
\begin{equation}
\label{singt}
t_{11}=t_{22}=t_{12} = {i\over \pi}\ln {b-a\over 4}\,\cdotp
\end{equation}
The equations (\ref{psising}) and (\ref{singt}) coincide nicely
with the corresponding equations (\ref{hol1f1c}) and (\ref{t11})
for the one-cut case.

The physical curves are characterized by the conditions (\ref{crgenW}) 
which read in the present case
\begin{equation}
\label{Wextr}
N_{1}t_{11} + N_{2}t_{12} \equiv k_{1}\ {\rm mod}\ N_{1}\, ,\quad
N_{2}t_{22} + N_{1}t_{12} \equiv k_{2}\ {\rm mod}\ N_{2}\, ,
\end{equation}
where $k_{1}$ and $k_{2}$ are integers. Those conditions are clearly
inconsistent with (\ref{tsing1}), which shows that physically an electric
cycle can never vanish (except of course if $N_{1}=0$ or $N_{2}=0$). On the
other hand, (\ref{singt}) implies that the magnetic cycle can vanish only
if
\begin{equation}
\label{condmags}
k_{1}\ {\rm mod}\ N_{1} \equiv k_{2}\ {\rm mod}\ N_{2}\, ,
\end{equation}
which is equivalent to the condition (\ref{magc}) used in the main text.
Equations (\ref{singt}) and (\ref{Wextr}) then show that the singular curve
satisfies the physical condition for the one-cut solution. This is an
important ingredient of the physical discussion in Section 3.

We now treat the particular example $N$ even and $N_{1}=N_{2}=N/2$ that is
relevant for section 3.2.2. There are $N^{2}/4$ vacua
$|k_{1},k_{2};N/2,N/2\rangle$ in the broken phase. Equation
(\ref{condmags}) can be satisfied only for the $N/2$ vacua
$|k,k;N/2,N/2\rangle$, and we focus on that case. The physical conditions
(\ref{Wextr}) then imply that $t_{11}-t_{22}\in\mathbb Z$, which means that
the two cuts play symmetric r\^oles. More precisely, the curve (\ref{RS})
must take the symmetric form
\begin{equation}
\label{symRS}
y^{2}= (z^{2} - a^{2})(z^{2}-b^{2})\, ,\quad z = x + {m\over 2g}\,\cvp
\end{equation}
where
\begin{equation}
\label{newv}
a_{1}=-{m\over 2g} - b\, ,\quad
b_{1} = -{m\over 2g} - a\, ,\quad a_{2} = -{m\over 2g} + a\, ,\quad
b_{2} = -{m\over 2g} + b\, .
\end{equation}
The general formulas (\ref{yeq}) and (\ref{R}) yield
\begin{equation}
\label{ggg}
a^{2}+b^{2}={m^{2}\over 2g^{2}}\,\cdotp
\end{equation}
The remaining physical condition reads
\begin{equation}
\label{remain}
e^{i\pi N (t_{11}+t_{12})} = 1\, .
\end{equation}
By using $t_{11}=t_{22}$ and $t_{12}=t_{21}$ it is straightforward to show 
that
\begin{equation}
\label{musymcase}
\mu_{1}+\mu_{2} = -2\oint_{\alpha}{x\, \d x\over y}\Biggm/
\oint_{\alpha}{\d x\over y}\,\cdotp
\end{equation}
Moreover, (\ref{symRS}) implies that
\begin{equation}
\label{simpl}
\oint_{\alpha} {x\,\d x\over y} = -{m\over 2g}
\oint_{\alpha} {\d x\over y}\,\cdotp
\end{equation}
The formula for $t_{11}+t_{12}$, which generically involves 
complicated elliptic integrals of the third kind, then
simplifies drastically,\footnote{I would like to thank V.~Kazakov for a 
discussion on this point.}
\begin{eqnarray}
t_{11}+t_{12} &=&{1\over 2i\pi}\lim_{\ell_{0}\rightarrow\infty}\Bigl(
\int_{\beta_{2}} {2x + \mu_{1} +\mu_{2}\over 
y}\,\d x - 4\ln\ell_{0}\Bigr)\nonumber\\
&=&  {1\over i\pi}\lim_{\ell_{0}\rightarrow\infty}\Bigl(
\int_{\beta_{2}} {z\,\d z\over y} - \ln\ell_{0}^{2}\Bigr)
= {i\over \pi} \ln {b^{2}-a^{2}\over 4}\,\cdotp\label{tau0si1}
\end{eqnarray}
The condition (\ref{remain}), together with (\ref{ggg}), finally implies
that the most general solution for the curve is given by (\ref{Swbb}).

\end{document}